\documentclass{aa}  

\usepackage{graphicx}
\usepackage{txfonts}
\usepackage{natbib} 
\bibpunct{(}{)}{;}{a}{}{,} % to follow the A&A style

\usepackage{verbatim} 
\usepackage{hyperref}   % Hyperlinks
\hypersetup{colorlinks=true,linkcolor=blue,citecolor=blue,filecolor=blue,urlcolor=blue}

\newcommand{\afe}{[\alpha/{\rm Fe}]}

\begin{document}

   \title{Impact of $\alpha$ enhancement on the asteroseismic age determination of field stars} 

   \subtitle{Application to the APO-K2 catalogue}
  \author{G. Valle \inst{1, 2}, M. Dell'Omodarme \inst{1}, P.G. Prada Moroni
        \inst{1,2}, S. Degl'Innocenti \inst{1,2} 
}
\titlerunning{APO-k2 teo}
\authorrunning{Valle, G. et al.}

\institute{
        Dipartimento di Fisica "Enrico Fermi'',
        Universit\`a di Pisa, Largo Pontecorvo 3, I-56127, Pisa, Italy
        \and
        INFN,
        Sezione di Pisa, Largo Pontecorvo 3, I-56127, Pisa, Italy
}

   \offprints{G. Valle, valle@df.unipi.it}

   \date{Received 13/12/2023; accepted 26/02/2024}

  \abstract
{}
{We investigated the theoretical biases affecting the asteroseismic grid-based estimates of stellar mass, radius, and age in the presence of a mismatch between the heavy element mixture of observed stars and stellar models. }
%%
% methods
{We performed a controlled simulation adopting a stellar effective temperature, [Fe/H], an average large frequency spacing, and a frequency of maximum oscillation power as observational constraints. Synthetic stars were sampled from grids of stellar models computed with different $\afe$ values from 0.0 to 0.4. The mass, radius, and age of these objects were then estimated by adopting a grid of models with a fixed $\afe$ value of $0.0$. The experiment was repeated assuming different sets of observational uncertainties. In the reference scenario, we adopted an uncertainty of 1.5\% in seismic parameters, 50 K in effective temperature, and 0.05 dex in [Fe/H]. A higher uncertainty in the atmospheric constraints was also adopted in order to explore the impact on the precision of the observations of the estimated stellar parameters.}
{Our Monte Carlo experiment showed that estimated parameters are biased up to 3\% in mass, 1.5\% in radius, and 4\% in age when the reference uncertainty scenario was adopted. These values correspond to 45\%, 48\%, and 16\% of the estimated uncertainty in the stellar parameters. These non-negligible biases in mass and radius disappear when adopting larger observational uncertainties because of the possibility of the fitting algorithm exploring a wider range of possible solutions. However, in this scenario, the age is significantly biased by $-8\%$.  
Finally, we verified that the stellar mass, radius, and age can be estimated with a high accuracy by adopting a grid with the incorrect value of $\afe$ if the 
metallicity [Fe/H] of the target is adjusted to match the $Z$ in the fitting grid. In this scenario, the maximum bias in the age was reduced to 1.5\%.}
{} 

   \maketitle

\section{Introduction}\label{sec:intro}

The chemical evolution of galaxies does not happen uniformly for the different elements \citep[see][and references therein]{Kobayashi2020}. In fact,
$\alpha$-process elements  (e.g., O, Mg, Ca, and Si) are
mainly synthesised in massive stars and core collapse
supernovae. In contrast, iron group elements such as Fe and Ni are mainly produced
by Type Ia supernovae, which involve less massive stars in binary systems. This introduces a time dependence of the relative abundance of the elements in the interstellar medium from which stars form.

A difference in the heavy element content proportions has a direct impact on the opacities of the stars, thus influencing their structure and the evolutionary timescale. The role played by $\alpha$ enhancement in stellar model computations is well studied and understood \citep[e.g.][]{VandenBerg2000, Kim2002, Pietrinferni2006, VandenBerg2014,  Fu2018, Pietrinferni2021}, and $\alpha$-enhanced stellar models have been widely used for different aims, such as estimating the ages of globular clusters \citep[e.g.][]{Lee2009, Cassisi2013, Gontcharov2019, Gontcharov2021}, studying elliptical galaxies \citep[e.g.][]{Thomas2003, Milone2007, Zhu2010}, estimating the age of stars in the Galactic bulge \citep{Bensby2013, bulge}, 
studying some of the oldest field stars in the Galaxy halo \citep{Grunblatt2021, Montalban2021}, and
obtaining asteroseismic estimates for stars in the Kepler catalogue \citep{Ge2015}.

Recently, $\alpha$-enhanced models have been adopted for the study of stars in the APOGEE-Kepler catalogue \citep{Pinsonneault2014}, which contains asteroseismic and spectroscopic data for thousands of stars in the red giant branch (RGB) \citep{Tayar2017, Salaris2018}. The relevance for Galactic archaeology investigations of a catalogue of RGB stars with precise asteroseismic, effective temperature, and metallicity observations has been widely recognised \citep[e.g.][]{Stello2015, Stasik2024}.  In fact, thanks to their high luminosity, RGB stars can be observed at larger distances than main-sequence (MS) stars. Moreover, RGB star oscillations can be investigated using a longer cadence than it is needed for MS stars.
The availability of this catalogue allowed \citet{Martig2015} to obtain grid-based estimates of mass, radius, and age for 1,639 stars in the RGB phase. Their investigation reported the puzzling presence of an unusual $\alpha$-rich young stellar population, confirming the detection by other surveys. The investigation was extended in \citet{Warfield2021}, who reported the existence of an intermediate age stellar population at high $\afe$. 

A great opportunity to gain further insight into the history of Galactic formation appeared with the publication of the APO-K2 catalogue \citep{Stasik2024}, which contains high-precision data for 7,673 RGB stars and combines spectroscopic \citep[APOGEE DR17;][]{Abdurrouf2022}, asteroseismic \citep[K2-GAP;][]{Stello2015}, and astrometric \citep[Gaia EDR3;][]{Gaia2021} data. This catalogue significantly improved the sampling of the region at high $\alpha$ enhancement.  
In \citet{Stasik2024}, mass and radius estimates were proposed for the stars in the catalogue by inverting the scaling relations that link these parameters to stellar effective temperature and two average asteroseismic observables, namely, 
 the large frequency spacing $\Delta \nu$ and the
frequency of maximum oscillation power $\nu_{\rm max}$. Masses and radii obtained by scaling relations are supposed to be robust and reliable, but they can be affected by a larger error than those from grid-based techniques \citep[see e.g.][]{Gai2011, Chaplin2014,  Pinsonneault2014, scepter1}. However, scaling relations cannot provide an estimate of the stellar age, which is obtained only by comparison with a grid of computed stellar models. 

A non-negligible problem that is often overlooked when performing a grid-based estimate of a fundamental stellar parameter is the assessment of its potential biases, which are expected to be different for different stellar evolutionary phases, observational uncertainties, and systematic errors \citep[e.g.][]{Gai2011, Basu2012, eta, Martig2015, smallsep}. In particular, the relevance of a systematic offset in the $\alpha$ enhancement observational constraint is still unexplored.  
In this paper, we focus on a theoretical investigation on the bias and the characterisation of the observational uncertainty propagation when adopting stellar grids with different values of $\alpha$ enhancement for stars mimicking those in the RGB phase in the APO-K2 catalogue. This preliminary work sets the stage for the grid-based analysis of the RGB stars in the APO-K2 catalogue (Valle et al., in preparation).

\section{Methods and stellar models grids}

\subsection{Stellar model grids}
\label{sec:grids}

Stellar models were computed using the FRANEC code \citep{scilla2008} in a configuration similar to that adopted to compute the Pisa Stellar
Evolution Data Base\footnote{\url{http://astro.df.unipi.it/stellar-models/}} 
for low-mass stars \citep{database2012}. 
The model grids were computed for masses in the range [0.75, 1.95] $M_{\sun}$, with a step of 0.01 $M_{\sun}$. The evolution was followed from the pre-main sequence until the RGB tip. 
Only models in the $\log g$ range [1.50, 3.25] with an age lower than 14 Gyr were retained in the grids.
The initial metallicity [Fe/H] was varied from $-1.5$ to 0.4 dex, with
a step of 0.025 dex. 
The solar heavy element mixture by \citet{AGSS09} was adopted. 
Five different values of $\alpha$ enhancement were allowed from 0.0 to 0.4 with a step of 0.1.
The initial helium abundance was fixed by adopting the commonly used
linear relation $Y = Y_p+\frac{\Delta Y}{\Delta Z} Z$
with the primordial abundance  $Y_p = 0.2471$ from \citet{Planck2020}
 and with a helium-to-metal enrichment ratio of $\Delta Y/\Delta Z = 2.0$ \citep{Tognelli2021}.
A solar-calibrated mixing-length parameter $\alpha_{\rm ml} = 2.02$ was adopted.  Outer boundary conditions were set by the \citet{Vernazza1981} solar semi-empirical $T(\tau)$, which approximates well the results obtained using the hydro-calibrated $T(\tau)$ \citep{Salaris2015, Salaris2018}.
Convective core overshooting was not included.
A moderate mass-loss was assumed according to the parametrisation by \citet{Reimers1975}, with the efficiency $\eta = 0.2$.
High-temperature ($T > 10,000$ K) radiative opacities were taken from the OPAL group
\citep{rogers1996},\footnote{\url{http://opalopacity.llnl.gov/}}
whereas for lower temperatures, the code adopts the molecular opacities by
\citet{ferg05}.\footnote{\url{https://www.wichita.edu/academics/fairmount_college_of_liberal_arts_and_sciences/physics/Research/opacity.php}} Both the high- and low-temperature opacity tables account for the metal distributions adopted in the computations. Each grid contains about six million points.
 
The average large frequency spacing $\Delta \nu$ and
the frequency of maximum 
oscillation power $\nu_{\rm max}$ were obtained using the scaling relations from
the solar values \citep{Ulrich1986, Kjeldsen1995} as follows:
\begin{eqnarray}\label{eq:dni}
        \frac{\Delta \nu}{\Delta \nu_{\sun}} & = &
        \sqrt{\frac{M/M_{\sun}}{(R/R_{\sun})^3}} \quad ,\\  \frac{\nu_{\rm
                        max}}{\nu_{\rm max, \sun}} & = & \frac{{M/M_{\sun}}}{ (R/R_{\sun})^2
                \sqrt{ T_{\rm eff}/T_{\rm eff, \sun}} }. \label{eq:nimax}
\end{eqnarray}
The validity of these scaling relations in the RGB phase has been questioned \citep[e.g.][]{Epstein2014, Gaulme2016, Viani2017, Brogaard2018, Buldgen2019}, and corrections accounting for the temperature and metallicity of the star have been proposed in the literature \citep[e.g.][]{Zinn2022, Stello2022}.
{ 
Although the paper explicitly neglects a potential systematic error arising from the incorrect adoption of asteroseismic parameters in the fitting process, this assumption may be overly optimistic when considering real stars. Discrepancies between the adopted asteroseismic parameters and the actual physical properties of the stars are to be expected. Notably, the metallicity dependence of the proposed correction to the large frequency separation makes it sensitive to inaccuracies in the adopted mixture model, as the conversion between Z and [Fe/H] is affected by such discrepancies. However, this effect is relatively minor, as we verified using the {\tt Asfgrid} code \citep{Sharma2016, Stello2022} on a sample of typical APO-K2 stars with $\afe \approx 0.3$, which yielded a median $\Delta \nu$ correction of merely 0.2\%, much smaller than the assumed observational errors.      
}

\subsection{Fitting pipeline}\label{sec:fittingML}

The fit was performed by means of the SCEPtER pipeline \citep{eta, bulge, binary, smallsep}. We briefly summarise the technique here for reader's convenience.

\begin{table}
        \centering
        \caption{Different observational uncertainties in the considered scenarios.} \label{tab:errors}
        \begin{tabular}{lcccc}
                \hline\hline    
                Scenario & $T_{\rm eff}$ (K) & [Fe/H] & $\Delta \nu$ & $\nu_{\rm max}$\\
                \hline
                E1 & 50 & 0.05 & 1.5\%  & 1.5\%\\
                E2 & 100 & 0.10 & 1.5\%  & 1.5\%\\
                E3 & 100 & 0.05 & 1.5\%  & 1.5\%\\
                \hline
        \end{tabular}   
\end{table}

We defined $q \equiv \{T_{\rm eff}, {\rm [Fe/H]}, \Delta \nu, \nu_{\rm max}\}$ as the vector of the observed quantities for a star and $\sigma$ as the vector of the corresponding observational uncertainties. We defined $\tilde q_j$ as the vector of observables for each point of the grid.
We computed the geometrical distance $d_{j}$ between the observed star and the $j$th grid point, defined as
\begin{equation}
        d_{j} = \left\lVert \frac{q - \tilde q_j}{\sigma} \right\rVert. \label{eq:dist}
\end{equation} 
The technique then computes the likelihood as
\begin{equation}
        L_j = \exp (-d_{j}^2/2).
        \label{eq:lik-scepter}
\end{equation}
This likelihood function was evaluated for each grid point within $2.5 \sigma$ of
all the variables from $q$, and we defined $L_{\rm max}$ as the maximum value
obtained in this step. The estimated stellar quantities were obtained
by averaging the corresponding mass, radii, and age of all the models with a likelihood
greater than $0.95 \times L_{\rm max}$.

Three different observational error sets were considered. The first set, E1, is the reference scenario. The errors in the asteroseismic quantities (1.5\%) were chosen to match the median uncertainty in the APO-K2 catalogue. However, the errors in effective temperature and [Fe/H] were increased with respect to those provided in APO-K2, that amount being about 7 K, arising from the reduction pipeline internal errors. In our work, we took into account that a much higher variability arises in the temperature and metallicity when different research is compared. We then adopted a cautious approach, assuming 50 K in $T_{\rm eff}$ and 0.05 dex in [Fe/H] as sensible uncertainties. The second scenario, E2, has increased uncertainty. This scenario considers a doubled uncertainty in effective temperature and metallicity, while the uncertainty in the seismic parameters is unchanged with respect to E1. The third scenario, E3, differs from E1 only because the effective temperature uncertainty is doubled. We only used this scenario in some specific tests. All the adopted uncertainties are summarised in Table~\ref{tab:errors}.

\section{Results}\label{sec:results}

\begin{table*}
        \centering
        \caption{Median and dispersion of the relative error in estimated masses, radii, and ages.} \label{tab:bias}
        \begin{tabular}{lccc|ccc|ccc}
                \hline\hline
                \multicolumn{10}{c}{Mass}\\
                \hline
                & \multicolumn{3}{c|}{E1} & \multicolumn{3}{c|}{E2} & \multicolumn{3}{c}{Equal $Z$}\\
                & $q_{16}$ & $q_{50}$ & $q_{84}$ & $q_{16}$ & $q_{50}$ & $q_{84}$ & $q_{16}$ & $q_{50}$ & $q_{84}$ \\ 
                \hline
                $[\alpha/{\rm Fe}] = 0.0$ & -6.9 & 0.0 & 6.8 & -7.3 & 0.0 & 7.0 &  &  &  \\ 
                $[\alpha/{\rm Fe}] = 0.1$ & -7.4 & -0.6 & 6.2 & -7.2 & 0.1 & 7.2 & -6.8 & 0.0 & 6.7 \\ 
                $[\alpha/{\rm Fe}] = 0.2$ & -8.1 & -1.3 & 5.4 & -7.0 & 0.3 & 7.5 & -6.8 & 0.0 & 6.9 \\ 
                $[\alpha/{\rm Fe}] = 0.3$ & -8.8 & -2.1 & 4.8 & -6.7 & 0.5 & 8.1 & -6.7 & 0.0 & 7.0 \\ 
                $[\alpha/{\rm Fe}] = 0.4$ & -10.2 & -3.2 & 4.2 & -6.7 & 0.7 & 8.4 & -6.6 & 0.1 & 7.0 \\ 
                Scaling relation & -7.3 & 0.0 & 7.9 & -7.7 & -0.1 & 8.3 & -7.4 & -0.1 & 7.9 \\ 
                \hline
                \multicolumn{10}{c}{Radius}\\
                \hline
                & \multicolumn{3}{c|}{E1} & \multicolumn{3}{c|}{E2} & \multicolumn{3}{c}{Equal $Z$}\\
                & $q_{16}$ & $q_{50}$ & $q_{84}$ & $q_{16}$ & $q_{50}$ & $q_{84}$ & $q_{16}$ & $q_{50}$ & $q_{84}$ \\ 
                \hline
                $[\alpha/{\rm Fe}] = 0.0$ & -3.1 & 0.0 & 3.0 & -3.3 & 0.0 & 3.0 &  &  &  \\ 
                $[\alpha/{\rm Fe}] = 0.1$ & -3.3 & -0.3 & 2.7 & -3.2 & 0.1 & 3.1 & -3.1 & 0.0 & 3.0 \\ 
                $[\alpha/{\rm Fe}] = 0.2$ & -3.7 & -0.6 & 2.3 & -3.2 & 0.0 & 3.1 & -3.0 & 0.1 & 3.0 \\ 
                $[\alpha/{\rm Fe}] = 0.3$ & -4.1 & -1.1 & 2.0 & -3.2 & 0.1 & 3.3 & -3.0 & 0.1 & 3.1 \\ 
                $[\alpha/{\rm Fe}] = 0.4$ & -4.8 & -1.5 & 1.6 & -3.2 & 0.0 & 3.4 & -3.0 & 0.1 & 3.1 \\ 
                Scaling relation & -3.4 & -0.1 & 3.4 & -3.5 & -0.1 & 3.4 & -3.5 & -0.1 & 3.4 \\ 
                \hline
                \multicolumn{10}{c}{Age}\\
                \hline
                & \multicolumn{3}{c|}{E1} & \multicolumn{3}{c|}{E2} & \multicolumn{3}{c}{Equal $Z$}\\
                & $q_{16}$ & $q_{50}$ & $q_{84}$ & $q_{16}$ & $q_{50}$ & $q_{84}$ & $q_{16}$ & $q_{50}$ & $q_{84}$ \\ 
                \hline
                $[\alpha/{\rm Fe}] = 0.0$ & -19.5 & 0.3 & 26.7 & -20.2 & 0.6 & 29.3 &  &  &  \\ 
                $[\alpha/{\rm Fe}] = 0.1$ & -20.1 & -0.1 & 26.4 & -23.1 & -2.6 & 25.7 & -19.6 & -0.2 & 25.7 \\ 
                $[\alpha/{\rm Fe}] = 0.2$ & -18.6 & 1.3 & 28.6 & -24.0 & -3.3 & 24.6 & -19.1 & 0.7 & 27.6 \\ 
                $[\alpha/{\rm Fe}] = 0.3$ & -18.5 & 2.6 & 30.8 & -27.0 & -5.6 & 21.7 & -19.3 & 1.2 & 27.9 \\ 
                $[\alpha/{\rm Fe}] = 0.4$ & -18.3 & 4.3 & 35.1 & -29.1 & -7.8 & 19.0 & -18.5 & 1.5 & 27.9 \\ 
                \hline
        \end{tabular}
        \tablefoot{The first column contains the $\afe$ value of the sampling grid (the fit was always performed using the $\afe  =0.0$ grid). Columns 2 to 4 contain -- for scenario E1 -- the median ($q_{50}$) and the 16th and 84th quantiles ($q_{16}$ and $q_{84}$) of the relative errors in mass, radius, and age. Columns 5 to 7 contain the same quantities for scenario E2. Columns 8 to 10 refer to the equal-$Z$ scenario. Values obtained from scaling relations are also reported.
        }
\end{table*}

We aim to quantify the biases and the uncertainty in the estimates of stellar age, mass, and radius due to the propagation of the observational uncertainty. Our main interest is in studying the trend of the biases when a mismatch exists between the $[\alpha/{\rm Fe}]$ values of the synthetic star and stellar models grids adopted in the fit. To this aim, we adopted the following procedure. In our step one, we created the artificial samples. We sampled at random $N = 10,000$ synthetic stars from each grid at $[\alpha/{\rm Fe}]$ from 0.0 to 0.4. Since the time step between consecutive points in the model grids is uniform, the sampling is not uniform in asteroseismic parameters. However, as we verified a posteriori, this does not bias the results.
Our step two concerned the perturbation of the stellar observables. Every sampled object was subjected to random Gaussian perturbations that mimicked the observational errors. This step was performed for the three chosen observational error sets: E1, E2, and E3 (Table~\ref{tab:errors}). Our third step was estimation. The mass, radius, and age of the artificial stars were then estimated by using the grid at $[\alpha/{\rm Fe}] = 0.0$, thus neglecting the information about the $\alpha$ enhancement. 
A comparison of the results obtained adopting different sets of observational uncertainties showed interesting features. 

\subsection{Estimated parameters in scenario E1}\label{sec:E1}

\begin{figure*}
        \centering
        \includegraphics[height=17.5cm,angle=-90]{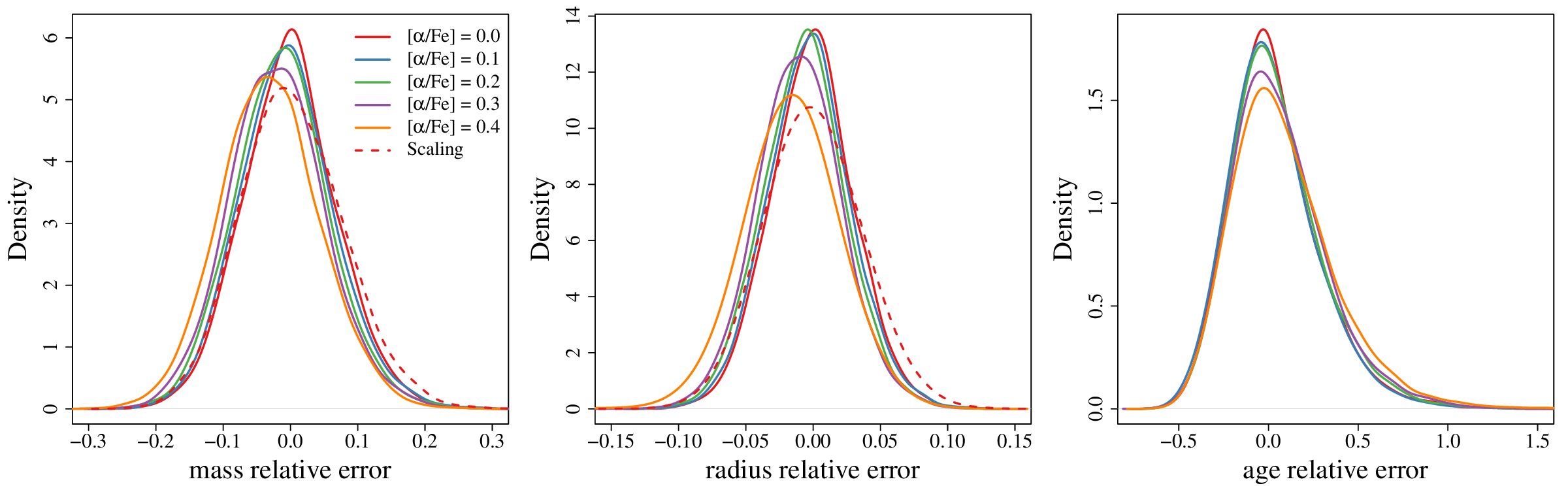} 
        \caption{Kernel density estimator of the relative errors in the estimated stellar parameters. {\it Left}: Mass relative error. The different colours codify the sampling grid, while the estimating grid is always that at $[\alpha/{\rm Fe}] = 0.0$. The dashed line corresponds to the estimation from the scaling relation.
        {\it Middle}: Same as in the left panel but for radius.
        {\it Right}: Same as in the left panel but for age.          
        }
        \label{fig:MRA-std}
\end{figure*}

The results obtained in the reference scenario are summarised in the first three columns in Table~\ref{tab:bias} and in Fig.~\ref{fig:MRA-std}.
The table presents the median values of the relative errors of the estimated quantities with respect to the true values (the values of mass, radius, and age of the sampled artificial object parameters). Given the possible non-normality of these samples, the median ($q_{50}$ in the table) was assumed as central estimator. To quantify the dispersion of the distributions, the 16th and 84th ($q_{16}$ and $q_{84}$) percentiles were adopted as a $1\sigma$ interval. Overall, the stellar parameters are estimated with noticeable precision whatever the sampling grid, with an average precision of about 7\% in mass and 3.5\% in radius. The age is less constrained, with a 25\% precision. 

\begin{figure*}
        \centering
        \includegraphics[height=17.5cm,angle=-90]{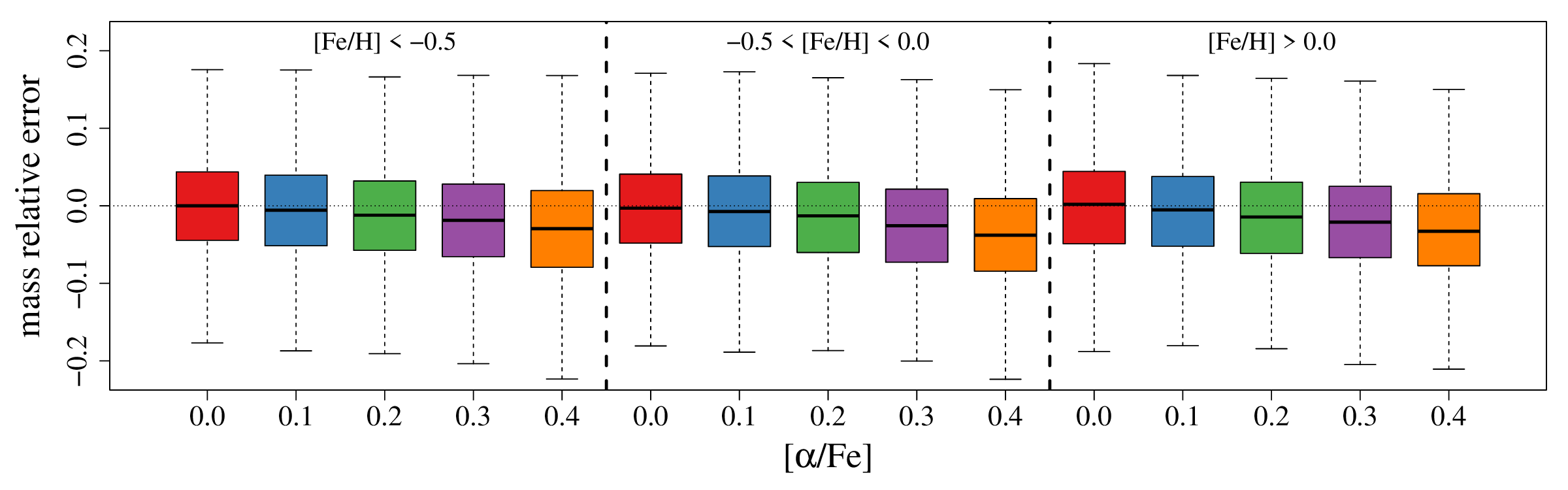} 
        \caption{Boxplot of the relative errors in mass according to different values $\afe$ of the sampling grid in three metallicity bins. The black line in the middle of the boxes marks the median, while the boxes cover the interquartile range. The whiskers extend to the extreme data. The colours correspond to different $\afe$ values in the sampling grids and match those in Fig.~\ref{fig:MRA-std}.     
        }
        \label{fig:M-vs-FeH}
\end{figure*}

The estimated mass bias increases with increasing mismatch between the $\afe$ in the recovering and in the sampling grids. The relevance of this bias when compared with the distribution dispersion also increases, raising from about 8\% for the sampling at $\afe = 0.1$ to about 45\% for  $\afe =0.4$. This bias trend is almost linear until $\afe = 0.3$, but it shows an increase at $\afe = 0.4$. The dispersion of the distributions is almost constant, with a 5\% increase for the sampling from the $\afe = 0.4$ grid. The reported trend and the anomaly at high $\afe$ have a simple explanation, which we discuss after presenting the results in the E2 scenario.
As a comparison, Table~\ref{tab:bias} also reports the results obtained by directly inverting the scaling relations in Eq.~(\ref{eq:dni}) and (\ref{eq:nimax}) to obtain the mass and radius of the artificial stars. As expected, the scaling relations are unbiased whatever the sampling grid. The table reports the average of them. It is interesting to note that the precision from the scaling relation (evaluated from the distribution dispersion) is comparable to that from the sampling from the unbiased grid $\afe = 0.0$, as it is only 10\% higher. This is not surprising because the constraints imposed by the grid morphology are less important in the RGB evolutionary phase than they are in earlier evolutionary stages. The narrow packing of the stellar tracks in RGB limits the relevance of the effective temperature and metallicity constraints \citep[see the analysis in][]{Valle2018}, with a consequent widening of the uncertainties in the grid-estimated parameters \citep[see e.g.][]{Gai2011, Pinsonneault2014, Martig2015, bulge, Moser2023}.
The detected biases do not depend on the metallicity, as shown in Fig-\ref{fig:M-vs-FeH}. This figure shows the relative error in the mass estimates, classified according to the sampling grid value of $\afe$, in three different metallicity intervals: [Fe/H] < -0.5; -0.5 < [Fe/H] < 0.0; and [Fe/H] > 0.0.  No relevant differences arose in the three metallicity zones.

Similar considerations also apply to our radius estimates. The relevance of the bias with respect to the dispersion increases from 9\% at $\afe = 0.1$ to 48\% at $\afe = 0.4$. The width of the distribution is about 6.1\% up to $\afe = 0.3$, and it slightly increases by 5\% at  $\afe = 0.4$. As with mass, the radius scaling relations have no bias either, with a dispersion about 10\% higher than the unbiased estimates from the sampling at $\afe = 0.0$.

Regarding age, the estimates show a positive bias, caused by the negative bias in mass estimates. Also in this case, the bias increases with increasing discrepancy between sampling and estimating grids $\afe$. However, given the higher uncertainty in the age estimates, the relative importance of the bias only gets to 16\% when stars are sampled from the $\afe =0.4$ grid.
Therefore, the age estimates are robust against a possible use in the fit of a grid adopting a wrong $\afe$  value.

\subsection{Estimated parameters in scenario E2}\label{sec:E2}

\begin{figure*}
        \centering
        \includegraphics[height=17.5cm,angle=-90]{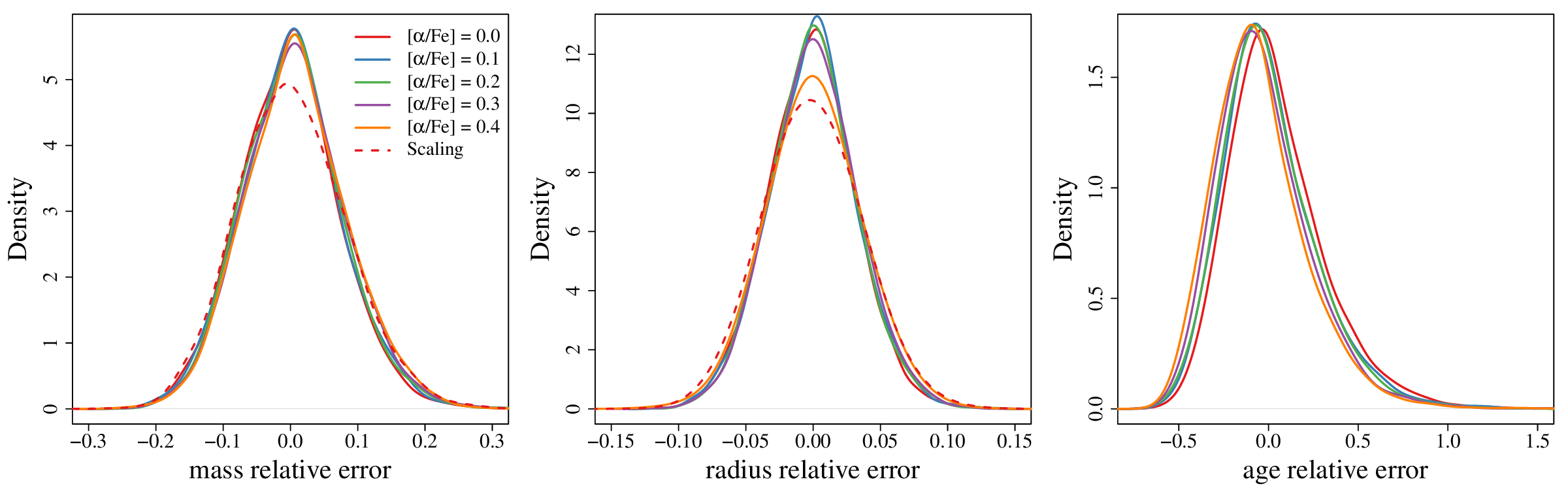} 
        \caption{Same as in Fig.~\ref{fig:MRA-std} but for a 100 K uncertainty in effective temperature and 0.1 dex in [Fe/H].    
        }
        \label{fig:MRA-E2}
\end{figure*}

\begin{figure*}
        \centering
        \includegraphics[height=17.5cm,angle=-90]{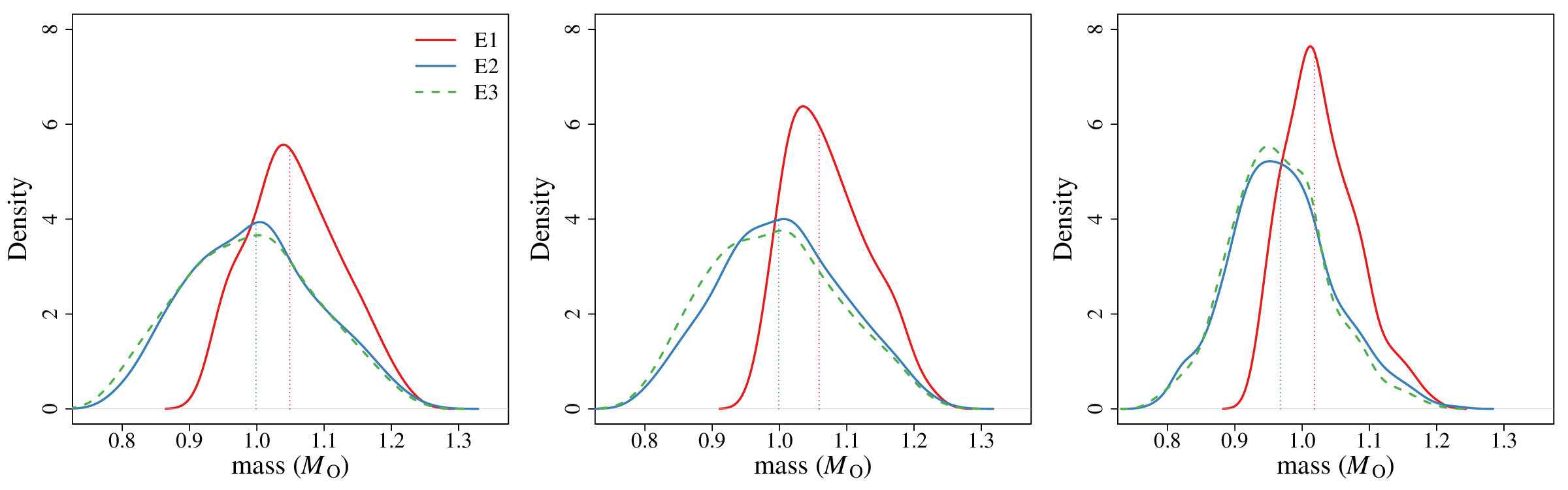} 
        \caption{Kernel density estimator of the mass distribution of models in the $[\alpha/{\rm Fe}] = 0.3$ grid entering the $2 \sigma$ box around the reference point at $M = 1.0$~$M_{\sun}$ and $\Delta \nu = 0.06$ in the $[\alpha/{\rm Fe}] = 0.0$ grid at three different initial [Fe/H] values.
                {\it Left}: Distribution of the masses at [Fe/H] = 0.2. The red line corresponds to error E1, the blue line to E2, and the green line to increased uncertainty only in effective temperature. The dotted lines mark the median of the respective distributions.
                {\it Middle}: Same as in the left panel but for [Fe/H] = 0.0.
                {\it Right}: Same as in the left panel but for [Fe/H] = -0.6.   
        }
        \label{fig:grid-mass}
\end{figure*}

The results obtained when adopting larger uncertainties in effective temperature and metallicity constraints shows interesting differences from those in scenario E1. The most apparent difference is the small or negligible biases in masses and radii (Tab.~\ref{tab:bias}) regardless of the sampling grid. The dispersion of the mass error distribution is only slightly higher than in scenario E1, which is not surprising because in the RGB, the fit variability is mainly dictated by the asteroseismic parameters \citep{Valle2018}. The maximum bias in mass, which occurs for $\afe = 0.4$, is only about 9\% of the distribution dispersion, one-fifth of the value in the E1 scenario. The situation is even more favourable for estimation of the radius, which appears to be obtained without bias for all the considered values of $\afe$. 

The difference with respect to the results in scenario E1 indeed has a simple explanation, as it is linked to the difference in effective temperature between the equivalent models (i.e. models with identical mass and [Fe/H] and in the same evolutionary phase) in grids with different values of $\afe$. The median difference in $T_{\rm eff}$ with respect to the grid at $\afe = 0.0$ is 13 K for $\afe = 0.1$, 34 K  for  $\afe = 0.2$, 53 K  for $\afe = 0.3$, and 76 K for $\afe = 0.4$. Scenario E1 assumes an uncertainty in effective temperature that is almost identical to the difference in $T_{\rm eff}$ between $\afe = 0.0$ and $\afe = 0.3$. As a consequence, given a model with mass $M_1$ from $\afe = 0.3$, its equivalent in the $\afe = 0.0$ grid is at about $1 \sigma$ far from it in the effective temperature. Therefore the neighbourhood of the synthetic stars will be populated by models with a mass distribution peaked away from $M_1$. The situation is even worse for data sampled from the $\afe = 0.4$ grid. 

In scenario E2, with its larger errors, the former difference reduces to about $0.5 \sigma$, allowing many more models to be considered in the neighbourhood of the synthetic star. The situation is shown in Fig.~\ref{fig:grid-mass}. In this figure, we adopt as a reference an $M = 1.0$ $M_{\sun}$ synthetic star from the $\afe = 0.0$ grid at $\Delta \nu/\Delta \nu_{\sun} = 0.06$,\footnote{As a reference, the points are before the RGB bump. For $\afe = 0.0$ its $\log g$ is about 2.8, while the RGB bump is at about $\log g = 2.55$.} for [Fe/H] = 0.2, 0.0, -0.6. For these three models, we constructed a $2 \sigma$ box around the synthetic star observables and plotted the distribution of the stellar masses selected in this way. It is apparent that these distributions are always biased when adopting E1 uncertainties, while the larger uncertainties in scenario E2 allow for an unbiased selection of masses. Obviously, this simple exercise does not tell the whole story because every model in the box should be weighted by its likelihood. Therefore, the large distortions in Fig.~\ref{fig:grid-mass} translate into much lower biases in the grid-estimated masses. As a comparison, the figure also reports the estimated mass distributions when adopting the errors of E3, that is, with the same effective temperature error as in E2 but the same [Fe/H] error as in E1. The distributions do not differ from those from scenario E2, confirming that the contribution of the uncertainty on [Fe/H] in RGB is of minor relevance for grid-based estimates \citep{Valle2018}. 
     
The unbiased estimates for the stellar mass come at the cost of a relevant bias in the age. The detected bias is about 8\% for $\afe = 0.4$, that is, about 32\% of the distribution dispersion. This bias is exactly what is expected when comparing the evolutionary time of equivalent models between the $\afe = 0.0$ and $\afe = 0.4$ grids.

\subsection{Estimated parameters imposing equal $Z$}\label{sec:EZ}

\begin{figure*}
        \centering
        \includegraphics[height=17.5cm,angle=-90]{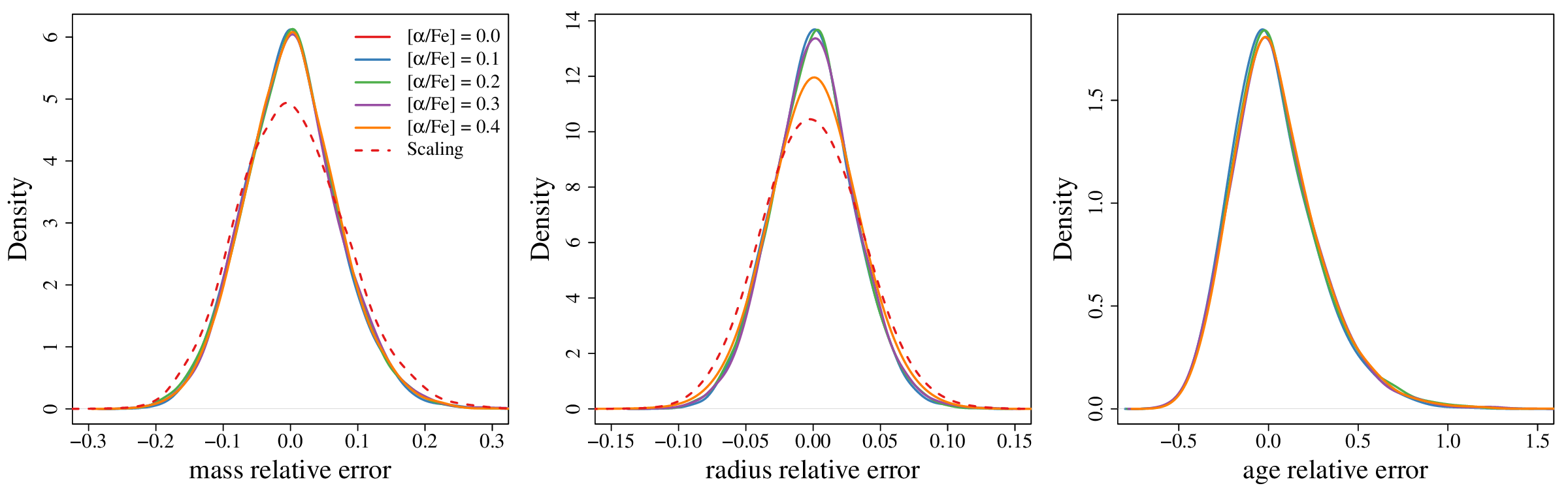} 
        \caption{Same as in Fig.~\ref{fig:MRA-std} but the surface metallicity of the $\alpha$-enhanced sampling grids has been adjusted to match the $Z$ value of the non-enhanced estimating grid.    
        }
        \label{fig:MRA-Z}
\end{figure*}

The lack of opacity tables for mixtures different from the solar one until the late 1990s posed a substantial problem to researchers investigating very metal-poor systems, such as globular clusters or elliptical galaxies, with  element abundance ratios different from
solar ones. This motivated a series of studies on the possibility of mimicking $\alpha$ enhancement by correcting the metallicity $Z$. Analytical formulae have been proposed in the literature for this purpose \citep{Chieffi1991, Chaboyer1992, salaris1993}.
These investigations have led to the common practice of mimicking $\alpha$-enhanced isochrones
by adopting more metal-rich non-enhanced isochrones, a practice still widespread in the recent literature \citep[e.g.][]{Martig2015, Warfield2021}.
However, it is well known that these simple analytical corrections cannot be blindly used whatever the [Fe/H] range, as they fail at high metallicity \citep[see e.g.][]{Kim2002}. It is therefore interesting to investigate how well a simple metallicity correction impacts the grid-based estimates.
In this section, we analyse the bias in stellar parameter estimates when adopting a similar approach.

\begin{figure*}
        \centering
        \includegraphics[height=17.5cm,angle=-90]{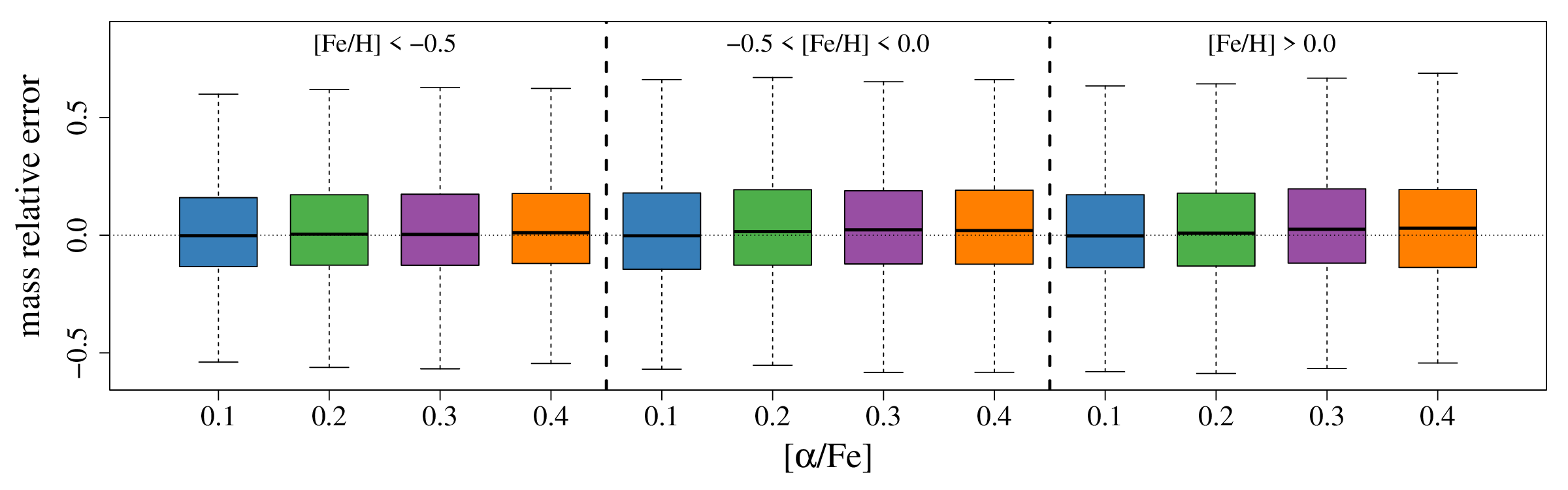} 
        \caption{Boxplot of the relative errors in age according to different values $\afe$ of the sampling grid in three [Fe/H] bins.   
        }
        \label{fig:MRA-Z-dep}
\end{figure*}

A straightforward way to correct for the $Z$ mismatch between $\alpha$-enhanced and non-enhanced models is to perform a correction in the [Fe/H] scale of the sampling grids. To this purpose, we computed the [Fe/H] for the $\alpha$-enhanced grids by matching the corresponding value on the non-enhanced grid at equal $Z$. These values were then adopted instead of the original [Fe/H] in the $\alpha$-enhanced grids. After this correction, the three-step procedure described at the beginning of Sect.~\ref{sec:results} was performed again. 
The results of the fit (Tab.~\ref{tab:bias} and Fig.~\ref{fig:MRA-Z}) showed a substantial reduction of the biases found in scenarios E1 and E2. The estimated mass of the artificial stars are always unbiased regardless of the $\afe$ of the sampling grid. The dispersion of the relative error in the mass estimates does not vary with $\afe$. The scaling relation estimates are about 10\% less precise than the grid-based ones. The same conclusions hold for the radius estimates.
The age estimates showed a uniform dispersion for different $\afe$ values of the sampling grid and a small residual bias that increases with $\afe$. This bias is higher at high values of [Fe/H] and disappears for [Fe/H] less than -0.5. As an example, for the $\afe = 0.4$ grid, the bias is about 1\% for [Fe/H] less than -0.5 and about 3\% for [Fe/H] greater than 0.0 (Fig.~\ref{fig:MRA-Z-dep}).
This result is not surprising because at high $Z$, the increased contribution of heavy elements to the total mass in the
mixture relative to hydrogen and helium causes a tiny difference in the evolutionary timescale ($\alpha$-enhanced models evolving faster) thus leading to the highlighted biases. As a comparison, the relative difference in the age of equivalent 1.0 $M_{\sun}$ models at $Z = 0.0017$ (corresponding to [Fe/H] = -0.9 in the $\afe = 0.0$ grid) between $\afe = 0.0$ and $\afe = 0.4$ grids is about 0.7\%, while it increases to about 2.6\% at $Z = 0.0301$ ([Fe/H] = 0.4 in the $\afe = 0.0$ grid).  
Overall, it appears that the usual correction based on matching $Z$ values is extremely effective when applied to grid-based estimates, achieving a performance nearly equivalent to the use of the $\afe$ unbiased grid in the fit of artificial stars.

\section{Conclusions}\label{sec:conclusions}

We investigated the bias affecting grid-based estimates when there is a mismatch in the adopted heavy element mixture between observed stars and the fitting grid of stellar models. We worked in a theoretical framework by computing stellar models with different values of $\afe \in [0.0, 0.4]$. Artificial stars, sampled from grids at various $\afe$ values, were subjected to estimations of mass, radius, and age, and we adopted a reference grid with $\afe = 0.0$. We adopted both asteroseismic ($\Delta \nu$ and $\nu_{\rm max}$) and classic ($T_{\rm eff}$ and [Fe/H]) observational constraints. We performed the estimation while adopting different 
sets of observational uncertainties. The uncertainty in asteroseismic quantities was set equal to the median value of what is found in the APO-K2 catalogue for RGB stars \citep{Stasik2024}, whereas different sets of uncertainty in $T_{\rm eff}$ and [Fe/H] were tested. We also investigated the common practice of  adjusting the target metallicity [Fe/H] to match the $Z$ value when $\alpha$ enhancement is neglected. The results of this paper will serve as a theoretical foundation for mass, radius, and age estimates (Valle et al. in preparation) of stars in the RGB phase in the APO-K2 catalogue.

The presented results show that the parameters of a target with known $\alpha$ enhancement can be estimated with high accuracy and precision with both a grid computed at the correct $\afe$ value as well as by adopting a grid with $\afe = 0.0$ and correcting the metallicity [Fe/H] of the target to match the $Z$ in the fitting grid. However, whenever the target $\afe$ is unknown and an incorrect grid is used, the estimated parameters are biased. This bias depends on two factors: The first factor is a mismatch between the $\afe$ of the target and that in the grid, which increases with the discrepancy in $\afe$. Given a difference in $\afe$ between a target and the grid, the bias also depends on the magnitude of the observational uncertainties, chiefly, that in the effective temperature. The bias is particularly relevant when the adopted uncertainty in $T_{\rm eff}$ is lower than the difference in effective temperature between equivalent stellar models in grids with two different $\afe$ values.

The detected biases and dispersion, however, should only be considered as representative of the random variability in the observations and due to the effect of the different levels of $\alpha$ enhancement. No systematic differences were allowed between the sampling and the fitting grids because their input physics are the same. Moreover, the adopted input physics may affect the results. We mention that a few key inputs may impact the results, including a difference in the helium-to-metal enrichment ratio; a difference in the efficiency of the microscopic diffusion, which affects the evolutionary timescale; or a difference in the adopted boundary conditions, with a consequent shift in the effective temperature of the RGB \citep[see e.g.][]{Salaris2018}.

\begin{acknowledgements}
G.V., P.G.P.M. and S.D. acknowledge INFN (Iniziativa specifica TAsP) and support from PRIN MIUR2022 Progetto "CHRONOS" (PI: S. Cassisi) finanziato dall'Unione Europea - Next Generation EU.
\end{acknowledgements}

\bibliographystyle{aa}
\bibliography{biblio}

\end{document}